\documentclass{mn2e}
\usepackage{amssymb}
\usepackage{ulem}
\usepackage{color}
\input epsf.sty
\newif\ifAMStwofonts
\AMStwofontstrue

\begin{document}

\title[QPO in AGN]
{Modeling the time-resolved quasi-periodic oscillations  in AGNs}

\author[Das \& Czerny]
  {Tapas K.~Das$^1$\thanks{tapas@mri.ernet.in} and 
B.~Czerny$^2$\thanks{bcz@camk.edu.pl}\\ 
  $^1$HRI, Chhatnag Road, Jhunsi, Allahabad 211 019, India\\
  $^2$Nicolaus Copernicus Astronomical Center, Bartycka 18, 
	00-716 Warsaw, Poland}

\maketitle
\begin{abstract}
Observation of the bright Seyfert 1 galaxy
RE~J1034+396 is believed to demonstrate a drift of the
central period of the Quasi Periodic Oscillation (QPO)
linearly correlated with the temporary X-ray luminosity. We show,
using a specific scenario of the oscillation mechanism in black hole 
accretion disc, that modeling such correlated trends puts very strong constraints on the nature
of this oscillation and the characteristic features of the hot flow in Active Galactic Nuclei (AGN). 
In our model, QPO oscillations are due to the oscillations of the shock formed in the low angular momentum hot 
accretion flow, and the variation of the shock location corresponds
to the observed changes in the QPO period 
and the X-ray flux. In this scenario, change in the shock location 
caused by perturbation of the flow 
angular momentum is compatible with the trends
observed in RE~J1034+396, whereas the perturbation of the specific flow energy results
in too strong flux response to the change of the oscillation period. Using a complete 
general relativistic
framework to study the accretion flow in the Kerr metric, we discuss the role of the black 
hole spin in the period drift. Future missions are expected to bring
more active galaxies with time-resolved quasi-periodic oscillations so similar
quantitative study for other QPO scenarios will be necessary.
\end{abstract}


\section{Introduction}

The details of the accretion flow pattern in AGN are far from being well understood.
One particular issue of profound interest in this regard is to predict the location of
the hard X-ray emitting region in such flows.
Spectral analysis of the present data is yet to provide any unanimously accepted
solution to this long standing question. Although the
presence of the relativistically smeared iron line imposes certain constraints 
(e.g. Fabian et al. 2002, Miniutti et al. 2007, 2009, Schmoll et al. 2009,
Niedzwiecki \& Zycki 2008, Niedzwiecki \& Miyakawa 2010; for a review, see e.g. Miller 2007),
the combined effects of the warm absorber and 
the reflection are, however, difficult to disentangle (for a review, see Turner \& Miller 2009). 

Future space missions, like the Astro-H, New Hard X-ray Mission (NHXM), International X-ray observatory (IXO), however,
will provide new opportunities to address
this issue. Several related proposals have already been put forward, e.g., the direct reverberation mapping
for the iron line (Armitage \& Reynolds 2003, Goosmann et al. 2007). Indications for some strong temporary
features have been, also for example, observed in the present X-ray data (see, e.g., Iwasawa et al. 2004, 
Ponti et al. 2004, Miller et al. 2006, Turner et al. 2006, Miniutti et al. 2007, Tombesi et al. 2007). 

In this work, we propose a novel theoretical approach to address this issue, 
based on the study of the expected properties of the QPO
in AGN. So far only one AGN showed indisputably 
a QPO episode
with a period of 3733 s (Gierlinski et al. 2008, Middleton et al. 2009), and it
corresponds to the 67 Hz QPO observed in GRS 1915+105 (Middleton et al. 2010). 
The detection was highly significant, and further study of the QPO time evolution during the single observation 
was possible with the use of the wavelet approach (Czerny et al. 2010).
A few other tentative detections have 
also been
suggested (e.g. Papadakis \& Lawrence 1993, Iwasawa et al. 1998, Fiore et al. 1989, 
Vaughan \& Uttley  2005, Lachowicz et al. 2006, Espaillat et al. 2008, 
Lachowicz et al. 2009), although most of them were later questioned as being only 
due to the
red noise (see Espaillat et al. 2008 for specific references).
However, further observations with current and future instruments will come up with high quality X-ray light curves of many more AGNs,
so even if the duty cycle for high-frequency QPO in AGN is as low as that observed for the galactic black hole
candidates, we expect to detect several QPO episodes (see e.g. Vaughan \& Uttley 2005), 
and the light curves will be of sufficiently high quality 
to study time-resolved frequency drift in a single QPO episode. The
signature of such a drift is believed to be already observed
using the XMM data for RE~J1034+396 (Czerny
et al. 2010), and the corresponding period modulation is accompanied by the proportional
change in the flux. In this work we show that the knowledge of the specific period-flux relation 
provides a good test of the QPO mechanism.

\section{Modeling the QPO}
\noindent
The characteristic features of the QPO
observed in the galactic and the extra galactic sources indicate
that such oscillations are a diagnostic of the accretion processes
in the inner region of the black hole accretion discs. Direct
analysis of the observational data shows that the high frequency 
QPO phenomenon appears only in a special spectral state - the steep
power law state, referred also as very high state, but the dynamical
QPO mechanism and the accretion flow pattern in this state are unknown. 
Several scenarios
of the QPO origin are under consideration. One of the possibility is that
QPO may be produced as a result of the oscillation of the shocks 
formed in hot accretion disc (Molteni, T\'oth \& Kuznetsov 1999,
Okuda, Teresi, Toscano \& Molteni 2004, Gerardi, Molteni \& Teresi 2005,
Okuda, Teresi \& Molteni 2007, and references therein). Such shocks may 
form when the angular momentum of the inflowing material is low, and the 
role of viscosity unimportant, while they are not expected in ADAF-type 
solutions (see e.g. Narayan et al 1997) where the angular momentum is only 
slightly sub-Keplerian and 
viscosity plays the dominant role in the inflow.

Based on the 
assumption of the fine tuning
of the inflow and the cooling parameters of a multi-transonic
black hole accretion, time dependent simulation work 
(see, e.g., Molteni, Sponholz \& Chakrabarti
1996) revealed that for a certain range of the initial boundary conditions, the 
radial oscillation of shock about their mean steady state location appear
and they 
cause the quasi periodic variation of the luminosity emerging out of the
post shock accretion flow (see also Giri et al. 2010 for more recent numerical
shock oscillation results). The frequency of such shock oscillation 
was shown roughly equal to the post shock advection time scale. This mechanism
may work if the whole inflow is optically thin or we have a two phase
medium, with the hot accreting corona  
dynamically decoupled from the underlying cold disk. 
We adopt and
test this scenario in our paper.

Since the post shock dynamical profile is essentially determined by the 
shock location (strength of the gravitational field) and shock 
compression (reduction of the dynamical velocity due to the shock),
such time scale effectively depends on some function of the shock location
$r_{sh}$ and the shock compression ration $R_{comp}$ as evaluated using 
the stationary solution (see, e.g., Das 2002, and references therein, for 
details of the calculation of $r_{sh}$ and $R_{comp}$ for a generalized  
multi transonic disc model). Hence from the theoretical front, the QPO 
frequency can reasonably be approximated by solving the stationary set of 
equations governing the axisymmetric flow of hydrodynamic multi-transonic
shocked accretion flow around astrophysical black holes (Chakrabarti 
\& Manickam 2000, Das 2003, Das, Rao \& Vadawale 2003, 
Mukhopadhyay, Ray, Dey \& Dey 2003, Mondal 2010). 
In this work, we follow Das (2003) and Das et al. (2003) to evaluate
the QPO frequency, which will be based
on the assumption that the low angular momentum hot inflow develops a shock, and the QPO
phenomenon represents the shock oscillation. To accomplish such task, one needs to self consistently
determine the location of the standing shock as well as the compression ratio
at the shock location. Both these variables are determined by three initial
boundary conditions, the specific flow energy ${\cal E}$ (also known as 
the Bernoulli's constant), the specific 
angular momentum $\lambda$ and the adiabatic index $\gamma$ of the flow  
(see, e.g., Das 2002 and Das, Bili\'c  \& Dasgupta 2007, and references therein, for the full details of such 
shock formation mechanism for the accretion flow under the influence of the pseudo-Newtonian 
black hole potential of Paczy\'nski and Wiita 
1980, and for full general relativistic flow analyzed in the Schwarzschild 
metric, respectively). 

One can also compute the QPO frequency in the full general relativistic 
framework, by considering the low angular momentum axisymmetric flow in 
the Kerr metric. $r_{sh}$ and $R_{comp}$ will then be determined by four 
parameters, ${\cal E},\lambda,\gamma$ and the Kerr parameter $a$, see,
e.g., Das \& Czerny 2009 and Barai, Chakraborty, Das \& Wiita 2009
for details of such calculations in the Kerr metric. 
Hence the most general shocked accretion flow, which will be used in this 
work to calculate QPO and the related quantities, are characterized by four 
parameters $\left[{\cal E},\lambda,\gamma,a\right]$. 

The aforementioned four parameters 
may further be classified intro three different groups, according to the
way they influence the fundamental features of the accretion flow. 
$\left[{\cal E},\lambda,\gamma\right]$ characterizes the {\it flow}, and not the
space time (since the accretion is assumed to be non self gravitating in our 
work, which is a common practice in the existing literature), whereas the Kerr
parameter $a$ exclusively determines the nature of the space time, and hence 
can be thought as some sort of `inner boundary condition' in qualitative 
sense (since the effect of gravity is determined within the full general 
relativistic framework only up to several gravitational radius -- beyond a certain
length scale, it asymptotically follows the Newtonian regime). Out of 
$\left[{\cal E},\lambda,\gamma\right]$, again, $\left[{\cal E},\lambda \right]$
determines the dynamical aspects of the flow, whereas $\gamma$ determines the 
thermodynamic properties. Hence to follow a holistic approach, one needs to 
study the variation of the observed phenomena on all of these four parameters.

If the parameters governing the flow vary slightly in time due to variable outer
boundary conditions the period of the 
QPO oscillation, as well as the overall X-ray emission will show a 
correlated drift in the similar pattern. We
parameterize such phenomenon by assuming a sequence of quasi-stationary solutions,
and quantify the characteristic trends of such a drift.
If the energy or the angular momentum of the flow are perturbed,
the position of the shock, and the corresponding frequency of the shock oscillation
and the radiation flux will then be altered. We calculate them in the following way.

The period of the oscillation is calculated from the formula
\begin{equation}
P_{QPO} \propto P_K(r_{sh}) R_{comp},
\label{eq:period}
\end{equation}
where $P_K(r_{sh})$ is the Keplerian period at the position of the shock location. The compression term increases the 
oscillation period since the oscillations are not strictly dynamic but due to the
coupled dynamical/thermal evolution, as shown (both analytically and numerically)
by Molteni et al. (1996).

The X-ray flux attributed to the shock may be approximated as
\begin{equation}
F \propto 2\pi r_{sh}H_{flow}c_s^2,
\label{eq:flux}
\end{equation}
where the sound speed, $c_s$, as well as the flow thickness
(disc height, which itself is a complicated function of the 
sound speed and other accretion parameters, see. e.g., 
Das 2002 and Das \& Czerny 2009 for the exact expression for the
$H_{flow}$ for pseudo-Newtonian and full general relativistic 
accretion flow, respectively) is measured in the post-shock configuration. 
The width of the shock
depends on the microscopic physics and we assume it is constant for all our models,
so we do not introduce it here as a specific additional parameter.

The numerical value of the proportionality constant for Eq. 
(\ref{eq:period}) and Eq. (\ref{eq:flux})
are unimportant since we are interested in testing the expected correlations between the radiation flux and the QPO central period 
in log-log space.
 Predicted trend can directly be
tested against the observed data.

\section{Results}

\begin{figure}
\epsfxsize=8.5cm
\epsfbox{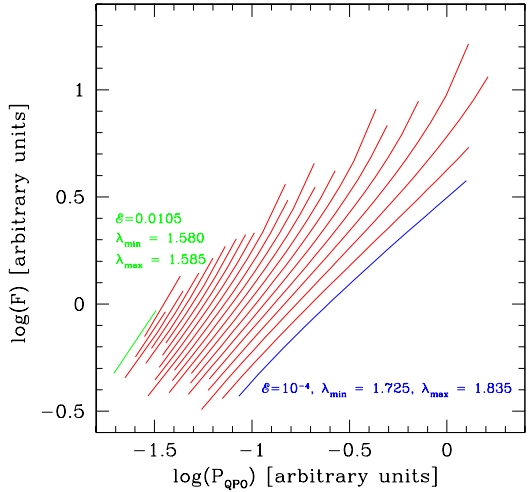}
\caption{The logarithmic plot (in arbitrary units) of the flux vs. period relation for a fixed 
specific energy $\cal E$, and variable angular momentum, $\lambda$, for a polytropic 
index $\gamma = 4/3$. Not all the 
parameter region is covered 
since shocks form only for a certain range of $\cal E$ 
and $\lambda$, for a given $\gamma$. The value of the 
specific energy as well as the maximum and the minimum value of the 
specific angular momentum for which the shock forms are 
shown for the leftmost (green line in the electronic version of the 
manuscript) and the rightmost (blue line in the electronic version of the 
manuscript) curve.}
\label{fig:flux_period1}
\end{figure}

Adopted scenario of the QPO mechanism in the form of shock oscillation in a hot axisymmetric 
low angular momentum flow allows us to connect the expected relation
between the X-ray flux and the central period of a QPO signal in a single AGN.

We first address the issue in the pseudo-Schwarzschild framework, by using 
the  Paczy\'nski and Wiita (Paczy\'nski and Wiita 1980) potential, and using 
the similar kind of flow geometry as described in Das, Rao \& Vadawale 2003. Hence the flow
is essentially governed by $\left[{\cal E},\lambda,\gamma\right]$ for this case.
We study the quasi-stationary response of the average flux to the forced change in the
QPO period. The stationary solution is specified by two dynamical
parameters (specific flow energy ${\cal E}$ and angular momentum $\lambda$),
which are determined by the
outer conditions in the flow which can vary with time.
The solution is characterized by the adiabatic index $\gamma$ as well, as described 
in \S 2.

We consider two types of perturbations corresponding to the aforementioned dynamical
parameters $\left[{\cal E},\lambda\right]$ separately. 

\begin{figure}
\epsfxsize=8.5cm
\epsfbox{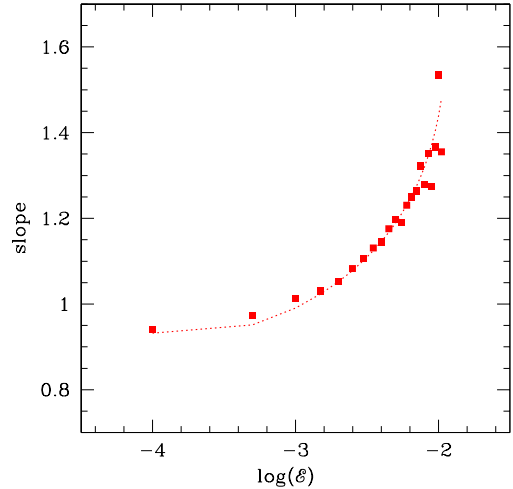}
\caption{The slopes of the logarithmic dependence of the  flux vs. period relation for a fixed specific
energy $\cal E$, and variable angular momentum, $\lambda$, for a polytropic 
index $\gamma = 4/3$. }
\label{fig:slopes_gamma43}
\end{figure}

First we fix the specific energy (i.e. the Bernoulli constant) and study the effect of the 
change of the angular
momentum of the flow. Such a perturbation can appear if the angular momentum of inflowing plasma 
at a given radius changes with time without any significant corresponding change in the flow 
temperature. The dependence 
of the flux on the QPO period is shown in 
Fig.~\ref{fig:flux_period1}. Each curve corresponds to a fixed Bernoulli constant, $\cal E$,
and the angular momentum varies along the curve, increasing from the 
left to the right of the figure.
The curves are almost straight lines in log plot, slightly
steepening at the longest period. However, the overall trend is well represented
with a straight line for shorter periods, so we calculate the line slope and give it
in Fig.~\ref{fig:slopes_gamma43}. The slope is flatter than 1 for small values of the Bernoulli
constant and rises considerably above 1 for $\cal E$ $>$ $ 3.0 \times 10^{-3}$. 

\begin{figure}
\epsfxsize=8.5cm
\epsfbox{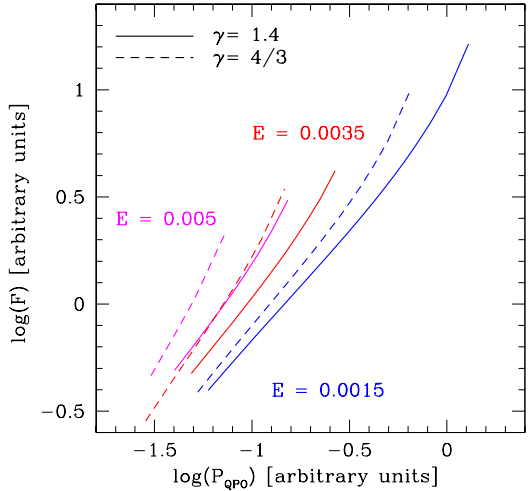}
\caption{The logarithmic plot of the flux vs. period relation for a fixed value of
specific energy
$\cal E$, and variable angular momentum, $\lambda$, for three pairs
of solutions for three different values of ${\cal E}$.
Each pair is characterized by two different values of the adiabatic 
indices, $\gamma = 4/3$ (marked by the dashed curve)
and $\gamma = 1.4$ (marked by the solid curve).}
\label{fig:flux_period2}
\end{figure}

The slope is only weakly sensitive on the adopted value of the parameter $\gamma$. If lower
value of the adiabatic index, e.g., $\gamma=1.4$, is used,
which is typically adopted for the partially ionized interstellar medium 
and for hot but magnetized ADAF (Advection-dominated Accretion Flow) plasma close to 
equipartition with the magnetic field (see, e.g. Esin 1997, 
Narayan et al. 1998), the slope becomes slightly stepper, particularly at high energies 
(see Fig.~\ref{fig:flux_period2}). Quantitatively,
the slope increases from 1.03 to 1.13 for $\cal E$$ = 1.5 \times 10^{-3}$, and from 1.20
 to 1.58 for  $\cal E$$ = 5 \times 10^{-3}$.

\begin{figure}
\epsfxsize=8.5cm
\epsfbox{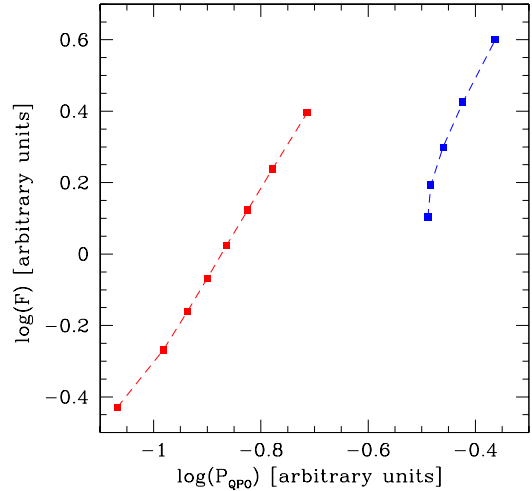}
\caption{The logarithmic plot of the flux vs. period relation for a fixed
angular momentum, $\lambda$, and variable asymptotic value of the Bernoulli constant,
$\cal E$. Two sequences represent two values of $\lambda$: 1.725 (left curve,
magenta coloured line in the online version) and 1.775 (right curve, blue coloured line
in the online version),
$\gamma = 4/3$.}
\label{fig:flux_period_lam1725}
\end{figure}

\begin{figure}
\epsfxsize=8.5cm
\epsfbox{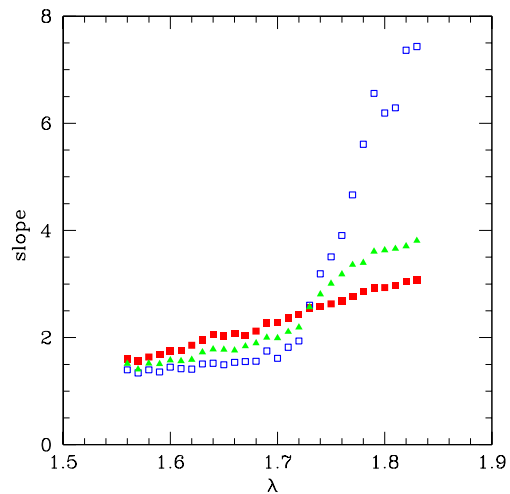}
\caption{The slopes of the logarithmic dependence of the  flux vs. period relation for a 
fixed angular momentum, $\lambda$, and variable Bernoulli
constant, $\cal E$, for a polytropic 
index $\gamma = 4/3$ measured in the linear part of the relation (red squares), in the high period  part (blue circles) and from the extreme points (green triangles). }
\label{fig:slopes_lam_43}
\end{figure}

We next study the case of the constant $\lambda$ by varying the asymptotic value of the 
Bernoulli constant, $\cal E$. This means that the angular momentum of the flow at a given 
radius remains constant while the plasma temperature falls or rises. In this case the 
change of the flux correlated with the change
of the QPO period is much more rapid (see Fig.~\ref{fig:flux_period_lam1725}). 
As before, we measure the slopes at the linear part of the log-log plot. The
slopes are much steeper, 2.29 for $\lambda = 1.725$, increasing even further with an
increase of $\lambda$. Thus the response of the flux is very sensitive to the 
nature and the characteristic features of perturbation involved. 
The range of the slopes of the flux-period relation as a function of the 
parameter $\lambda$ is shown in Figure~\ref{fig:slopes_lam_43} as the set of red squares.
Since in the case of constant $\lambda$ the curvature is more significant than in the case of 
constant $\cal E$ perturbations, in addition we also show slopes measured in the upper parts 
of every $log F - log P_{QPO}$ curve (blue circles) and the slopes just measures as a difference 
between the two extreme points on the curve (green triangles). 

The two types of the perturbations thus differ significantly in the predicted slope in the 
flux-period relation. Perturbations which involve both the change of the specific flow energy
and the angular momentum should show intermediate slopes between the two extreme cases.

\begin{figure}
\epsfxsize=10.5cm
\epsfbox{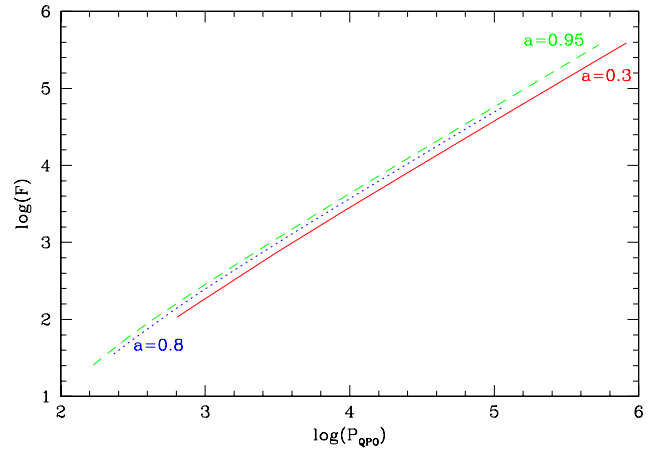}
\caption{For complete general relativistic flow in the 
Kerr metric, the logarithmic plot of the flux vs. period relation for a fixed Bernoulli
constant, ${\cal E}=1.0000044$, and variable angular momentum, $\lambda$ is shown.
Three sequences represent three values of the Kerr parameter, $a=0.3,0.8$ and $a=0.95$
respectively.
The value of the adiabatic constant is taken to be $\gamma = 4/3$.}
\label{fig:flux_period_kerr}
\end{figure}

Finally, we examined whether the results depend significantly on the spin of the
black hole. Our
basic model (Das et al. 2003) was developed for the 
Paczy\'nski and Wiita (Paczy\'nski and Wiita 1980) pseudo-Schwarzschild potential, well 
describing a non-rotating black hole. Therefore, to study the effect of the Kerr metric, we 
used more general flow equations, as described in detail by Das \& Czerny (Das \& Czerny 2009)
which allows to
determine the shock location and the flow parameters
within a complete general relativistic frame work. We then use Eqs.~\ref{eq:period}
and \ref{eq:flux} taking local flow quantities without any
relativistic corrections for the outflowing X-ray flux. Three examples of the flux-period
relation are shown in Fig.~\ref{fig:flux_period_kerr}. The slope of this relation 
does not change with the Kerr parameter because the shock does not form very close to the
horizon where the black hole spin has leading effects. The representative value of the specific 
flow energy (general relativistic Bernoulli 
constant),including the rest mass factor, is taken to be ${\cal E}=1.0000044$
(similar figures can also be drawn for any other required value of ${\cal E}$), which is
close to the lowest value used to construct Fig.~\ref{fig:slopes_gamma43}, hence the direct
comparison is possible. Of course the range of the frequencies has changed
but this is relevant only if there is an accurate determination of the black 
hole mass involved in the process. The effect is due to the fact that the local properties
of the flow do not strongly depend on the Kerr parameter at distances of $\sim 10 R_g$,
where the shock forms, and the flow parameters are measured. 
However, the global properties of the solutions do depend strongly on the Kerr parameter,
as discussed in Barai, Chakraborty, Das \& Wiita 2009 (see, e.g., figure 5 and 6 of of this paper).

The model does not have any explicit dependence on the accretion rate, and does not
imply any such constraints on inflowing material. This is valid as long as the accretion rate, and the 
flow optical depth are not too high. Our model does not include the effect of the radiative cooling 
on the flow dynamics, and this assumption breaks down for accretion rates of the hot inflow 
above a fraction of  the Eddington rate, when the optical depth for the electron scattering is $\sim 1$.
 
\section{Discussion}

We study a shock oscillation scenario of the QPO oscillations suggested by 
Molteni et al. (1996) and
later considered by a number of authors as described in the section 2.
Within the framework of this model, thermal/dynamical oscillations of the shock 
correspond to the QPO frequency, and deviations from the shock location due to the
change of the asymptotic values of the flow result in a correlated change in the
QPO period as well as in the flux averaged over QPO period.
The well specified model allows us to predict the slope of this correlated change
of the period and the flux, which can be compared with the value 
directly obtained
from the X-ray data.

There are also other mechanisms proposed for explanation of QPO: 
(i) Trapped pulsating
modes with an inner disk/torus (e.g. Perez et al. 1997, Wagoner et al. 2001, Espaillat et al. 2008, 
Straub \& Sramkova 2009), (ii) Temporary spots on
the disk surface (e.g. Karas 1999, Schnittman 2005, Bachetti et al. 2010), (iii) Epicyclic coupled oscillations in both vertical and 
radial directions 
from exact planar motions within a disk (e.g., Abramowicz \& Kluzniak 2001, 
Abramowicz 2005, Horak et al. 2009; 
see also Stella \& Vietri 1999), 
and (iv) Disk-jet magnetic coupling (e.g. Wang et al. 2007). Some of those models,
like oscillations of the radiating tori, have in principle the 
required predictive power to give the expected flux-period relation but the computations
are time consuming so single examples only are shown in the respective works. In other
cases, like epicyclic coupled oscillations, the models do not contain yet the predictions
of the X-ray emissivity. Shock oscillations can already be tested, although under
some assumptions, and this issue will be addressed in Sect.~\ref{sect:shock_apli}.

\subsection{Application to RE~J1034+396}

The QPO phenomenon was discovered in this source by Gierlinski et al. (2008). More 
detailed analysis of the same XMM data sequence by Czerny et al. (2010) based
on wavelet analysis indicates that the QPO period shows a drift, which was accompanied 
by a change of the X-ray flux, with the index of the power law trend of $0.92 \pm 0.03$.
If this value is indeed representative for the QPO in this source, then it is
consistent with predictions of the shock model in case of purely angular momentum 
perturbations, and the Bernoulli constant of order of $\cal E$$ \sim 10^{-4}$ or lower.

Such perturbations seem to be more likely in the case of a turbulence in a magnetized plasma; magnetorotational instability (MRI; Balbus \& Hawley 1991) in such a flow may lead to angular momentum redistribution without very significant heating/cooling of the plasma. However, at this moment no adequate simulations of the process are available. The MHD computations of the low angular momentum flow were performed mostly in 2-D approximation (Proga \& Begelman 2003, Okuda et al. 2007, Moscibrodzka \& Proga 2009), 3-D computations were done without the magnetic field (Janiuk et al. 2009), only Gerardi et al. (2005) performed 3-D MHD, but all those papers neglected dissipation. Some 3-D MHD do include radiative losses (e.g. Hirose, Krolik \& Stone 2006, with subsequent applications) but those computations assume shearing box and almost Keplerian disk.  

The energy constraint can be translated to the hot plasma temperature at the outer boundary, 
and the implied temperature is $\sim 300$ keV or lower, i.e. comparable to the electron 
temperature of the plasma considered in the coronal models of the origin of the X-ray 
radiation.

Both constraints are consistent with the likely global scenario of the accretion inflow in this object. The broad
band spectrum of REJ1034+396 is dominated by the multi-color black body component (Puchnarewicz et al. 2001, Loska et
al. 2004, Middleton et al. 2009), and additional the power low component is steep (Middleton et al. 2009). We
identify this power law emission with low angular momentum accretion flow discussed in detail in the present paper.
The fraction of energy emitted by the disk is much higher so  the accretion goes prenominantly through this channel,
and the accretion rate in the hot plasma is only a small fration of the total accretion rate. The temperature of the
plasma responsible for this power law emission cannot be determined observationally because the high energy cut-off
is not seen, but the value of 300 keV mentioned above is thus consistent with the data. However, in the future the
model should be further developed to include the radiative coupling between the two phases and to test the possible
role of the dynamical (magnetic) coupling of the cold and hot flow using some simple parametric approach. 

\subsection{Prospects for further QPO detections in AGN with current and future experiments}

As discussed by Vaughan \& Uttley (2005), the exact estimate of the future QPO detections
are not quite straightforward. Simple increase in the detector area will not help 
for a specific source if the QPO frequency happens to be in a frequency range dominated
by the red noise. However, the actual detection of the QPO in RE~J1034+396 shows
that the detection is viable even now, with the XMM-Newton telescope, for a typical 
duration of an observation. The QPO with the period of $\sim 1 $ hour detected in  RE~J1034+396 implies the expected periods in other AGN of similar order, or somehat longer, if the black hole mass is larger. However, even shorter periods may be expected since the RE~J1034+396 is an equivalent to
67 Hz QPO in GRS 1915+105 (Middleton et al. 2010), and in some other AGN we may see the frequences corresponding to kHz QPO typical for other galactic sources. The only problem is that the number of lightcurves of
this S/N quality is small, and with the expected short duty cycle, confirmed by the
absence of the QPO in the first part of the 2007 data as well as in whole 2009 data
(C. Done, P. Lachowicz, private communications), the probability to pinpoint a source in the 
right state is low. Another source may show a QPO phenomenon in the XMM-Newton or Suzaku 
observation any time,
but telescopes with
bigger area like the proposed IXO mission would bring good S/N lightcurves for numerous AGN. Also 
other missions with smaller detector area but focused on extension of the detection towards
harder X-rays may bring QPO phenomenon more effectively, since the QPO signal is more profound
at higher energies (Middleton et al. 2009).  

\subsection{Applicability of the shock model to disk-dominated states}
\label{sect:shock_apli}

The dynamical model of the flow relies on the assumption of constant, low angular momentum.
If the accretion flow is dominated by the standard disk, the model still applies to the
coronal hot flow if the disk/corona coupling is negligible. Such sandwich disk/corona
flow has been considered by several authors (e.g. Meyer \& Meyer-Hofmeister 1994, 
\. Zycki et al. 1995, Witt et al. 1997, Rozanska \& Czerny 2000, Meyer et al. 2000, Mayer \&  Pringle 2007,
Kawanaka et al. 2008, Liu \& Taam 2009).
On the other hand, the inner flow is likely
to be dominated by the hot flow in the QPO state since no variations in the cold
disk are seen in QPO spectra of galactic sources (e.g. Sobolewska \& \. Zycki 2006). 
Also QPO in RE~J1034+396 are related entirely to the hot flow (Middleton et al. 2009). 
The assumption of bremstrahlung cooling in the simulations performed by Molteni 
et al. (1996) has further been relaxed in recent hydrodynamical simulations performed by Giri et al. (2010)
where it has been demonstrated that the oscillations appear even in the adiabatic case, essentially due to a 
possibility of a temporary backflow. In this case the radiation is a small perturbation 
to the dynamics of the flow and outgoing spectrum can be dominated
by Comptonization, as observed in reality.

Calculation of the flow spectra is beyond the current model, as this would require determination of the electron temperature of the flow, and additional assumptions on the dissipation within the shock. We can only state that there is no basic discrepancy between our model and the observed spectrum fitted by Middleton et al. (2009). The fraction of luminosity in the hard power law tail comming from the hot flow is small, the exact value strongly depends on the adopted model as the decomposition of the X-ray spectrum into soft and hard components is not unique, and the hard X-ray slope changes with the model change from $\Gamma = 2.3$ to $\Gamma = 3.6$. The high energy extension of this tail is also not determined so the optical depth of the hard X-ray emitter is unknown. 

\subsection{Applicability of the study to Galactic Black Holes}

Accretion processes onto black holes show amazing similarity across the very broad range
of black hole masses from $\sim 10 M_{\odot}$ Galactic sources to $10^{10} M_{\odot}$
bright quasars. However, the correlations studies here are not expected to apply to
galactic sources under the present time resolution. The observed correlation
obtained for RE~J1034+396 showed a period trend within a {\it resolved} single QPO
episode, and the model presented here also aims to simulate such a situation. 
This QPO oscillation may be an analog of the 67 Hz high frequency QPO in GRS 1915+105
(Middleton \& Done 2010). High 
frequency QPOs in Galactic sources are, at present, unresolved within their dynamical
timescale, so the observed 
correlations, if any,
would imply secular trends corresponding to a major rearrangement of the flow 
geometry in a long visccous timescale. Some trends are actually seen in accreting
neutron star systems, but then the increase in the period corresponds to the decrease
in the overall emissivity, which is in opposite direction than the trend discussed here. However,
if the future instrumental time resolution and the photon countrate allow to see the
evolution of a period during a single QPO episode, the expected correlations is likely to apply
for such case. 

\section*{Acknowledgments}

This work was supported in part by N N203 380136.
The visit of BCZ at HRI was partially supported by astrophysics project
under the XIth plan at HRI. Useful comments made by an anonymous 
referee is acknowledged.

\ \\
This paper has been processed by the authors using the Blackwell
Scientific Publications \LaTeX\ style file.

\end{document}